\documentclass[journal]{IEEEtran}
\usepackage{cite}
\usepackage{times}
\usepackage{amsmath}
\usepackage{epsfig}
\usepackage{caption}
\usepackage{graphicx}
\usepackage{epstopdf}
\usepackage{subfigure}
\usepackage{multirow}
\usepackage{siunitx}
\usepackage{bm}
\usepackage{csquotes}
\usepackage{braket}
\usepackage{color}

\allowdisplaybreaks
\newcommand{\figref}[1]{{Fig.~\ref{#1}}}

\ifCLASSINFOpdf

\else

\fi

\begin{document}
%
\title{Analysis of the interaction between classical and quantum plasmons via FDTD-TDDFT method}

\author
{
 Jian Wei~You,~\IEEEmembership{Member,~IEEE,}
and~Nicolae C.~Panoiu,~\IEEEmembership{Member,~IEEE}
\thanks {This work was supported by the European Research Council (ERC), Grant Agreement no. ERC-2014-CoG-648328.}
\thanks{J. W. You and N. C. Panoiu are with the Department of Electronic and Electrical Engineering, University College London, London, WC1E7JE, UK. (e-mail: j.you@ucl.ac.uk and n.panoiu@ucl.ac.uk)}
\thanks{Manuscript received December 15, 2018.}
}

\markboth{IEEE Journal on Multiscale and Multiphysics Computational Techniques,~Vol.~XX, No.~XX, XX~2019}%
{Shell \MakeLowercase{\textit{et al.}}: Analysis of the interaction between classical and quantum plasmons via FDTD-TDDFT method}

\maketitle

\begin{abstract}
A powerful hybrid FDTD--TDDFT method is used to study the interaction between classical plasmons
of a gold bowtie nanoantenna and quantum plasmons of graphene nanoflakes (GNFs) placed in the
narrow gap of the nanoantenna. Due to the hot-spot plasmon of the bowtie nanoantenna, the
local-field intensity in the gap increases significantly, so that the optical response of the GNF
is dramatically enhanced. To study this interaction between classical and quantum plasmons, we
decompose this multiscale and multiphysics system into two computational regions, a classical and a
quantum one. In the quantum region, the quantum plasmons of the GNF are studied using the TDDFT
method, whereas the FDTD method is used to investigate the classical plasmons of the bowtie
nanoantenna. Our analysis shows that in this hybrid system the quantum plasmon response of a
molecular-scale GNF can be enhanced by more than two orders of magnitude, when the frequencies of
the quantum and classical plasmons are the same. This finding can be particularly useful for
applications to molecular sensors and quantum optics.
\end{abstract}


\IEEEpeerreviewmaketitle

\section{Introduction}

\IEEEPARstart{T}{he} interaction of light with metallic nanoparticles has been a central theme in plasmonics
\cite{sam07book,pb13book,vgf16NanoPhot} ever since the beginnings of this discipline. In
particular, electromagnetic waves can induce collective oscillations of free electrons in metals,
giving rise to so-called plasmon modes. In recent years, plasmons in two-dimensional (2D)
materials, such as graphene, have attracted increasing research interest, primarily because of the
new rich physics characterizing these materials. For instance, since plasmons are confined in a
very small region, the induced optical near-field in these low-dimensional materials can be
enhanced and localized in extreme-subwavelength regions \cite{vgf16NanoPhot,sbc10NatMat}, which can
be used, \textit{e. g.} to plasmon-enhanced spectroscopy \cite{xbk99PRL} and light concentration
beyond diffraction limit \cite{gb10NatPhot,tmo13NatPhys}. Moreover, these properties are dependent
on the material structure and geometrical configuration of the plasmonic system, its size, and the
electromagnetic properties of the surrounding medium \cite{wv07ARPC,skd12Nature}, thus they can be
readily tuned. As a result, plasmonic effects have found a plethora of applications, including the
design of highly integrated nanophotonic systems \cite{sbc10NatMat,gb10NatPhot,po04nl,mis04PRL},
nanophotonic lasers and amplifiers \cite{tmo13NatPhys,bl12NatPhot}, new optical metamaterials
\cite{hpm12NatMat}, nanoantennas \cite{xbk99PRL,gfh11CR}, single-molecule spectroscopy
\cite{zzd13Nature}, photovoltaic devices \cite{po07ol,cc14NatPhot}, surface-enhanced Raman
scattering \cite{xbk99PRL,jbm03JPCB}, higher-harmonic light generation
\cite{fzp06nl,kjk08Nature,kkf10NanoLett,psl18jo}, catalytic monitoring of reactant adsorption
\cite{llz09Science}, sensing of electron charge-transfer events \cite{nfm08NatNano}, and biosensing
\cite{ahl08NatMat,rlj15Science}.

When the geometrical size of plasmonic nanoparticles is less than about \SI{10}{\nano\meter}, the
description of their optical properties becomes more challenging because quantum effects begin to
play an important role. At this scale, plasmon resonances become more sensitive to the quantum
nature of the conduction electrons \cite{skd12Nature}, thus the theoretical predictions of
approaches based entirely on the Maxwell equations are less successful in describing experimental
results \cite{she12Nature,rkc15NatComm}. The shortcomings of the classical theory stem chiefly from
neglecting three quantum effects: \textit{i}) spill-out of electrons at medium boundaries
\cite{zeb16NatComm,tsk15NatComm}, \textit{ii}) surface-enabled electron-hole pair creation
\cite{yg2008ss,lxz2013njp,cyj17PRL}, and \textit{iii}) nonlocal effects of electron wavefunction
\cite{lga17Science,mrw14NatComm}. These quantum effects can significantly change the features of
plasmon spectra predicted by the classical theory \cite{ywm15PRL,km15ACSPhot,myd17jpcc}. For
example, the spill-out effect results in an inhomogeneous permittivity around the nanoparticle
boundary, a phenomenon responsible for the size-dependent frequency shift of plasmon resonances
\cite{vgf16NanoPhot,skd12Nature,rkc15NatComm,bm17NanoPhot}. In order to overcome these shortcomings
of the classical theory, a new research area that combines plasmonics with quantum mechanics, known
as quantum plasmonics, has recently emerged
\cite{vgf16NanoPhot,tmo13NatPhys,zeb16NatComm,bm17NanoPhot}.

Theoretical and experimental advances in plasmonics have been greatly facilitated by the
development of efficient numerical methods. In the classical regime, the physical properties of
plasmonic nanostructures can be modeled numerically by solving the Maxwell equations using specific
computational electromagnetic methods, such as the finite-difference time-domain (FDTD) method
\cite{taf05book}. In these computational methods, the classical plasmon features are mainly derived
from the specific geometrical configuration of the plasmonic system and the distribution of the
(local) dielectric function. As long as the size of a plasmonic system is large enough, its
electromagnetic response is very well described by this computational classical approach. However,
if the size of nanoparticles is smaller than about \SI{10}{\nano\meter}, quantum effects must be
included in the computational description of plasmonic nanostructures. One such numerical method,
widely used in quantum plasmonics, is the time-dependent density functional theory (TDDFT)
\cite{ull11book,mmng12book}. In particular, it has been used to describe the dynamic response of
plasmonic systems and the corresponding optical spectra \cite{km15ACSPhot,mkn12NanoLett}. However,
TDDFT calculations are very time and memory consuming, thus most TDDFT simulations are
limited to small nanoparticles (generally, less than a few hundreds atoms)
\cite{yyg07PRL,zpn09NanoLett}.

Driven by the need to describe the physical regime in which classical and quantum effects play
comparable roles, there has recently been growing interest in developing classical-quantum hybrid
methods that can address this regime. For example, several numerical methods that aim to describe
strongly-coupled classical-quantum hybrid systems were proposed
\cite{man11nanolett,dfbg14JO,srn14JPCM,pc16JPCC,sn17JPCM} and used to investigate phenomena such as
super-radiance, Rabi splitting, and nonlinear harmonic generation. However, in order to optimize
the computational costs and simplify the numerical analysis, some recent works
\cite{cmrs10JPCC,cmrs12JPCC,sll15PCCP} demonstrated that in specific situations it is enough to
only consider the direct coupling of the classical system to the quantum one and neglect the back
coupling (the so-called weak-coupling regime). Here, we should note such decoupled analysis assumption is only applicable in the weak-coupling regime. In the strong-coupling regime, full-quantum methods or two-way-coupling classical-quantum simulations should be used.

One such physical system is studied in our paper, namely we apply an FDTD--TDDFT hybrid numerical
method \cite{cmrs10JPCC} to study the classical-quantum plasmon interaction in the weak-coupling
regime. In particular, we investigate the interaction between a classical plasmon of a bowtie
nanoantenna and a quantum plasmon of a graphene nanoflake (GNF) in the weak-coupling limit. In this
multiphysics and multiscale approach, the classical optical response of the bowtie nanoantenna is
computed using the FDTD method, whereas the quantum optical properties of the molecular-scale GNF
are determined using the TDDFT. Importantly, we employ the local field calculated with the FDTD
method as excitation field used in the quantum mechanical calculations, thus bridging the
descriptions of the classical and quantum components of the hybrid plasmonic nanosystem.

The remainder of this paper is organized as follows. In Sec.~\ref{sec:PhyComp}, we present the multiphysical system investigated in this
study, and the corresponding computational methods are presented as well. In Sec.~\ref{sec:ResDis}, we first describe the optical
properties of the plasmons of the GNF and bowtie nanoantenna, respectively, and then the main features of the interaction between these two types of plasmons is presented. Finally, the main conclusions are summarized in the last section.

\section{Multiphysical system and computational methods}\label{sec:PhyComp}
\begin{figure}[t!]
\centering\includegraphics[width=9 cm]{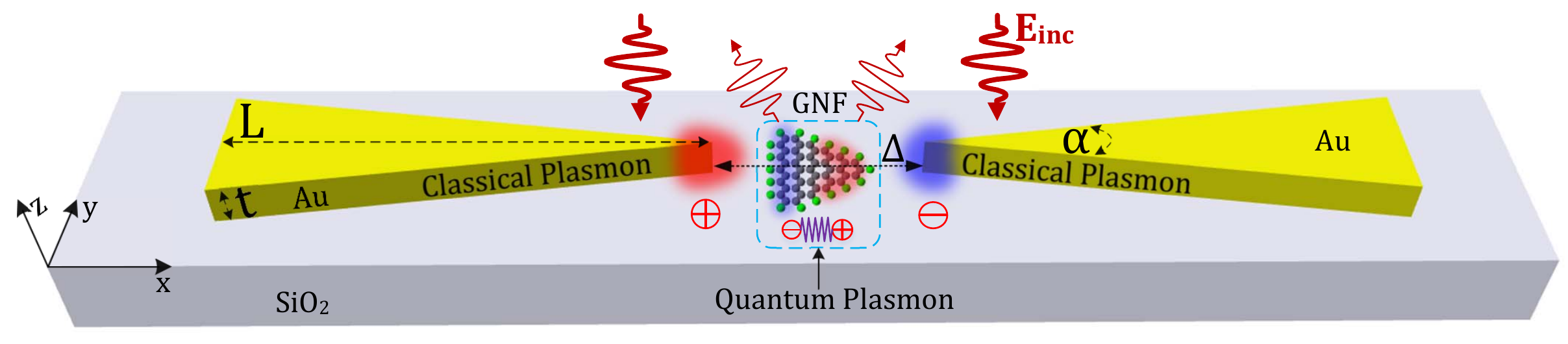}
\caption{Multiphysical system of a
classical and quantum hybrid plasmon model, where the quantum plasmon of a graphene nanoflake
interacts with the classical plasmon of a gold bowtie nanoantenna \textit{via} the optical
near-field.} \label{fig:PhyMech}
\end{figure}

The physical system used to illustrate the main features of our numerical method is shown in
\figref{fig:PhyMech}. It consists of a gold bowtie nanoantenna placed on a silica substrate and a
molecular-scale GNF located in the narrow gap of the nanoantenna. The nanoantenna is made of two
triangular gold plates with angle, $\alpha$, length, $L$, and thickness, $t$, the separation
distance between the tips of the gold plates being $\Delta$. By properly choosing the computational
grid, we made sure that the edges of the two triangular gold plates were flat. In all our
simulations $\alpha=\ang{12}$, $\Delta=\SI{10}{\nano\meter}$, and $t=\SI{30}{\nano\meter}$, but $L$
will be varied. Moreover, we assume that the GNF has a triangular shape, too, with side length,
$a=\SI{1.23}{\nano\meter}$, namely there are six carbon atoms along each side of the triangle. It
should be noted that triangular GNFs is one of the stable configurations in which they exist
\cite{sb11bCh}. The GNF is positioned in such a way that its symmetry axis coincides with the
longitudinal axis of the nanoantenna.

The bowtie nanoantenna has plasmon resonances associated with the triangular plates and strongly
localized (hot-spot) plasmons generated in the narrow gap of the nanoantenna. Given the size of the
bowtie nanoantenna, the spectral characteristics of these plasmons can be determined by solving the
Maxwell equations. The physics of the molecular-scale GNF, on the other hand, must be described
using quantum mechanics based numerical methods. Due to the large size difference between the two
plasmonics systems, there is a very large mismatch between the computational grids on which the
classical and quantum dynamics are resolved, as well as the corresponding time scales. Before we
describe the hybrid numerical method that allows one to overcome these difficulties, we will
briefly present the FDTD and TDDFT methods used in the classical and quantum computations,
respectively.

In order to numerically solve Maxwell's equations using the FDTD method \cite{taf05book}, one
should discretize them on the Yee grid. As a result of this procedure, one obtains the following
system of iterative equations:
\begin{equation}
\begin{aligned}\label{eq:IterativeE}
\left. {{E_d}} \right|_{i,j,k}^{n + 1} = {\alpha _1}\left. {{E_d}} \right|_{i,j,k}^n + {\alpha
_2}\left( {\left. {\Delta {l_{d - 1}}{{\bar H}_{d - 1}}} \right|_{i,j,k}^{n + \frac{1}{2}} }\right. \\
\left.{ - \left. {\Delta {l_{d + 1}}{{\bar H}_{d + 1}}} \right|_{i,j,k}^{n + \frac{1}{2}} - \left. {\Delta s_{d}{J_{s,d}}} \right|_{i,j,k}^{n + \frac{1}{2}}} \right)
\end{aligned}
\end{equation}

\begin{equation}
\begin{aligned}\label{eq:IterativeH}
\left. {{H_d}} \right|_{i,j,k}^{n + \frac{1}{2}} = {\beta _1}\left. {{H_d}} \right|_{i,j,k}^{n -\frac{1}{2}} + {\beta _2}\left( { \left. {\Delta {l_{d + 1}}{{\bar E}_{d + 1}}} \right|_{i,j,k}^n }\right. \\
\left.{ -\left. {\Delta {l_{d - 1}}{{\bar E}_{d - 1}}} \right|_{i,j,k}^n - \left. {\Delta s_{d}{M_{s,d}}}
\right|_{i,j,k}^n} \right)
\end{aligned}
\end{equation}

Here, the subscripts $i$, $j$, and $k$ indicate the grid points on the Yee grid, and $d=x$, $y$,
and $z$ indicates the field component. The subscript $d$ is used in a circular way, \textit{i.e.}
if $d=x$, then $d-1=z$ and $d+1=y$. The area of each face of the Yee grid is calculated as $\Delta
s_{d} = \Delta l_{d-1}\Delta l_{d+1}$, where $\Delta l_{d}$ is the length of the corresponding
edge. Moreover, the time step is indicated by an integer $n$, and the spatial difference operator
is represented by $\bar E$ and $\bar H$, which are defined as follows:
\begin{equation}\label{eq:Ebar}
\left. {{{\bar E}_{d \pm 1}}} \right|_{i,j,k}^n = \left. {{E_{d \pm 1}}} \right|_{{{\left\langle
{i,j,k} \right\rangle }_{d \mp 1}} + 1}^n - \left. {{E_{d \pm 1}}} \right|_{i,j,k}^n
\end{equation}
\begin{equation}\label{eq:Hbar}
\left.{{{\bar H}_{d \pm 1}}} \right|_{i,j,k}^n = \left. {{H_{d \pm 1}}} \right|_{{{\left\langle {i,j,k}
\right\rangle }_{d \mp 1}} + 1}^n - \left. {{H_{d \pm 1}}} \right|_{i,j,k}^n
\end{equation}

Here, the notation $<i,j,k>_{d \pm 1}+1$ indicates a shift of the grid index ($i$,$j$,$k$) with
respect to the subscript $d \pm 1$. For instance, when $d=x$, we have $d-1=z$, thus the notation
$<i,j,k>_{d-1}+1=<i,j,k>_z+1=(i,j,k+1)$. Moreover, the iteration coefficients in
Eqs.~\eqref{eq:IterativeE} and \eqref{eq:IterativeH} are
\begin{equation}\label{eq:Alpha1}
{\alpha_1} = (2\varepsilon -\sigma_{e} \Delta t)/(2\varepsilon + \sigma_{e} \Delta t)
\end{equation}
\begin{equation}\label{eq:Alpha2}
{\alpha_2} = {2\Delta t}/{\left[({2\varepsilon + \sigma_{e} \Delta t})\Delta s \right]}
\end{equation}
\begin{equation}\label{eq:Beta1}
{\beta_1} = (2\mu -\sigma_m \Delta t)/(2\mu + \sigma_m \Delta t)
\end{equation}
\begin{equation}\label{eq:Beta2}
{\beta_2} = {2\Delta t}/{\left[({2\mu  + \sigma_m \Delta t})\Delta s \right]}
\end{equation}
where $\Delta t$ is the time step, $\varepsilon$ is the electric permittivity, $\mu$ is the magnetic permeability, $\sigma_{e}$ is the electric
conductivity, $\sigma_m$ is the equivalent magnetic loss.

Based on the formalism described above, an FDTD iteration is performed by repeating the following
three steps until a preset convergence criterion is satisfied \cite{ytc14MTT}: In \textit{Step 1},
using the spatial distribution of the fields $H$ and $E$ at the time steps
$n-\frac{1}{2}$ and $n$, respectively, one updates the $H$ field at the time step
$n+\frac{1}{2}$ \textit{via} Eq.~\eqref{eq:IterativeH}. In \textit{Step 2}, using the spatial
distribution of the fields $H$ and $E$ at the time steps $n+\frac{1}{2}$ and $n$,
respectively, one updates the $E$ field at the time step $n+1$ \textit{via}
Eq.~\eqref{eq:IterativeE}. Finally, at \textit{Step 3}, one sets $n=n+1$ in \textit{Step 1} and
repeats \textit{Step 1} through \textit{Step 3}. At the end of each iteration one verifies whether
the preset convergence criterion is satisfied.

The quantum mechanical calculations are based on the TDDFT method, which we briefly present in
what follows. One starts from the Schr\"{o}dinger equation,
\begin{equation}\label{eq:TDSE}
i\hbar \frac{\partial }{\partial t} \ket{\Psi(t)} = \hat{H}\ket{\Psi(t)},
\end{equation}
where $\hat{H}$ is the Hamiltonian of the system, $\ket{\Psi(t)}$ is the many-body wavefunction,
and $\hbar$ is the reduced Planck constant. In the general case, the Hamiltonian of the system can
be written as:
\begin{equation}\label{eq:HamiltOp}
\hat H = {\hat T_n} + {\hat T_e} + {\hat V_{nn}} + {\hat V_{ee}} + {\hat V_{ne}} + {\hat V_{s}},
\end{equation}
where $\hat{T}_n$ and $\hat{T}_e$ are the kinetic energy of the nuclei and electrons, respectively,
$\hat{V}_{nn}$ and $\hat{V}_{ee}$ are the nuclei-nuclei and electron-electron Coulomb interactions,
respectively, $\hat{V}_{ne}$ is the nuclei-electron potential, and $\hat{V}_s$ is the external
potential, \textit{e.g.} the interaction potential with an applied external electric field (for
more details, see \cite{ull11book,mmng12book}).

For a general many-body system, the complexity of Eq.~\eqref{eq:TDSE} makes it very challenging to
solve it directly, especially for systems with large number of electrons. In order to overcome this
challenge, one usually employs two approximations: \textit{i}) One assumes that the properties of
atoms are mainly determined by the valence electrons, and \textit{ii}) the ionic cores are assumed
to be fixed. Moreover, using the Hohenberg-Kohn \cite{HK64PR} and Runge-Grosss \cite{rg84PRL}
theorems, the many-body wavefunction is expressed as a Slater determinant formed with
single-particle orbitals of a noninteracting system, $\psi_{i}(\mathbf{r},t)$, whose dynamics are
determined by the effective Kohn-Sham Hamiltonian, $\hat{H}_{KS}$:
\begin{equation}\label{eq:TDKS}
i\hbar \frac{\partial }{\partial t}\psi_{i}(\mathbf{r},t) =
\hat{H}_{KS}(\mathbf{r},t)\psi_{i}(\mathbf{r},t).
\end{equation}
In this formalism, the effective Kohn-Sham Hamiltonian is expressed as follows:
\begin{align}\label{eq:HamiltOpKS}
\hat{H}_{KS}(\mathbf{r},t) =& \hat{T}_{e}(\mathbf{r},t) + \hat{V}_{H}[\rho(\mathbf{r},t)] + \hat{V}_{xc}[\rho(\mathbf{r},t)] \nonumber \\
& + \hat{V}_{ne}(\mathbf{r},t) + \hat{V}_{s}(\mathbf{r},t),
\end{align}
where the Hartree potential, $\hat{V}_{H}[\rho(\mathbf{r},t)]$, which represents the classical
Coulomb electron-electron interaction and the exchange-correlation potential,
$\hat{V}_{xc}[\rho(\mathbf{r},t)]$, are functionals that depend on the electron charge density,
\begin{equation}\label{eq:ChargDens}
\rho(\mathbf{r},t)=\sum_{i}\vert\psi_{i}(\mathbf{r},t)\vert^{2}
\end{equation}
It should be noted that the
exchange-correlation potential ${V_{xc}}[{\rho({{\bf{r}},t})}]$ is generally unknown, so that in
practice several approximations of various degrees of sophistication are used, including the local
density approximation \cite{ks65PR} and the generalized gradient approximation \cite{pcvj92PRB}.

In summary, a self-consistent TDDFT iteration consists of four steps. \textit{Step 1}: The
ground-state wavefunction is calculated using the DFT method and then is used to construct the
initial electron density, $\rho(\mathbf{r},t_{k})$. \textit{Step 2}: The Hamiltonian $\hat{H}_{KS}$
defined by Eq.~\eqref{eq:HamiltOpKS}, corresponding to the time step $t_k$, is constructed using
the electron density $\rho(\mathbf{r},t_{k})$. \textit{Step 3}: Using the Hamiltonian
$\hat{H}_{KS}$ and Kohn-Sham orbitals $\psi_i$ calculated at $t_{k}$, the orbitals $\psi_i$ and the
electron charge density $\rho(\mathbf{r},t_{k+1})$ at the time step $t_{k+1}$ are calculated using
a proper time propagator \cite{cmr04JCP}. \textit{Step 4}: test the convergence criterion. If
convergence is reached, the iteration is stopped, otherwise the iteration is repeated from
\textit{Step 2}.

The two computational methods, FDTD and TDDFT, are decoupled and can be used independently of each
other. There is, however, a way to use them together when one aims to describe physical systems
that contain both classical and quantum components. We illustrate this approach on our plasmonic
system. Thus, we use first an FDTD simulation to determine the optical spectrum of the gold bowtie
nanoantenna and the time dependence of the electric field at the location of the GNF. Then, in a
subsequent simulation, this electric field is used as excitation field (external potential) in an
TDDFT simulation and the corresponding optical spectra of the GNF are computed. This hybrid
computational approach rigourously takes into account the influence of the classical plasmon of the
bowtie nanoantenna on the quantum plasmon of the GNF and the accompanying field enhancement
effects, but it does not incorporate the back-coupling effect of the quantum plasmon on the
classical one. It is expected that this effect is very small considering the size mismatch between
the classical and quantum objects, which would mean that our method provides reliable predictions.

\begin{figure}[!t]
  \centering
  \includegraphics[width=9 cm]{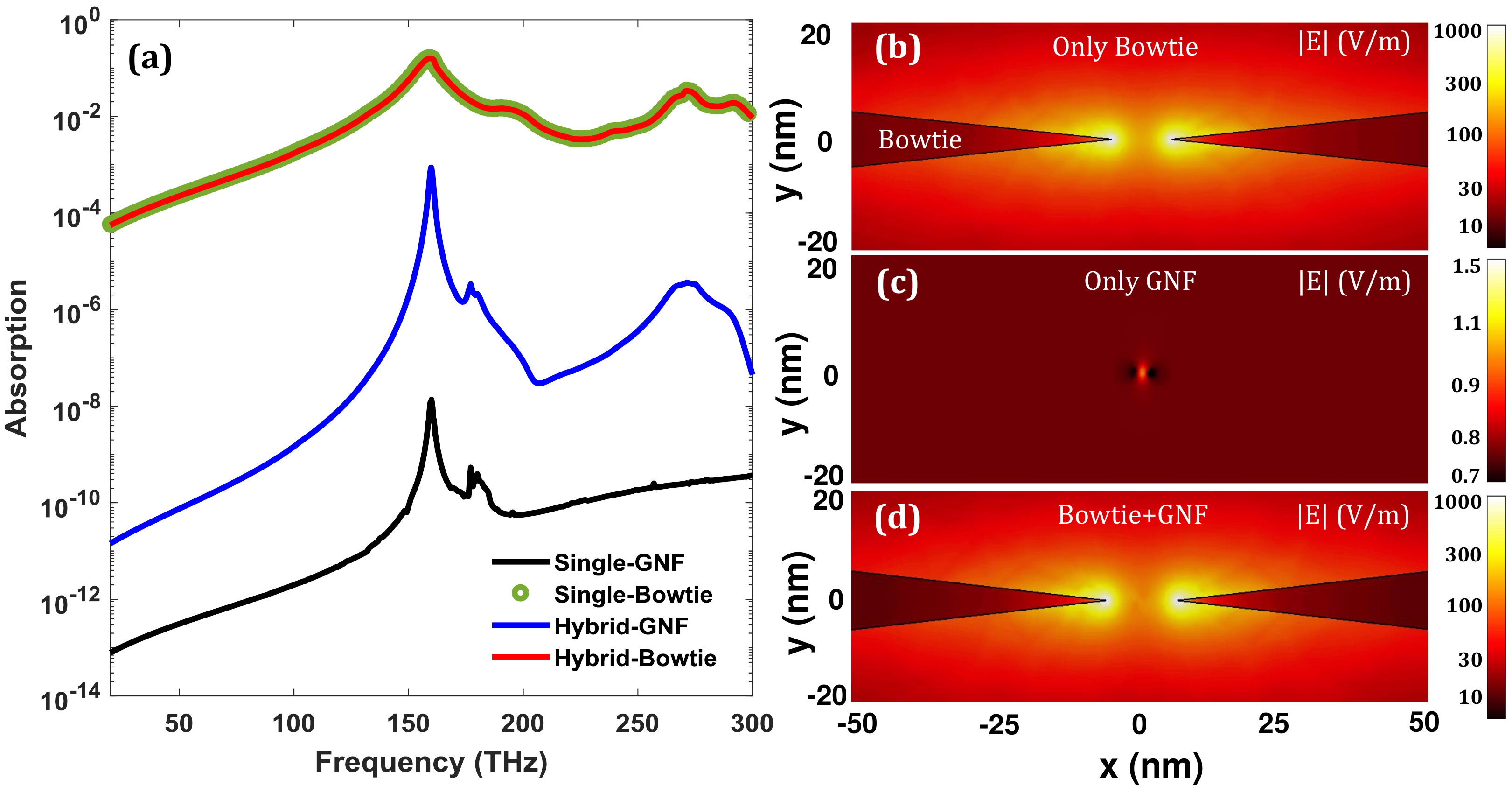}
\caption{Classical simulation. (a) Absorption of an
isolated GNF, an isolated bowtie antenna, GNF in a bowtie-GNF hybrid system, and bowtie in a
bowtie-GNF hybrid system. The absorption in the bowtie nanoantenna corresponds to the domains shown
in the right panels and not the entire nanoantenna. (b), (c), (d) Field-distributions of an
isolated bowtie nanoantenna, an isolated GNF, and a bowtie-GNF hybrid system, respectively.}
\label{fig:PhyMechAnalog}
\end{figure}

In order to validate this key assumption, we have calculated the absorption spectra of an isolated
gold bowtie nanoantenna, an isolated GNF, and a bowtie-GNF classical system, with the results of
this analysis being presented in \figref{fig:PhyMechAnalog}. In these calculations, the length of
bowtie antenna was $L=\SI{280}{\nano\meter}$ and the side length of triangular GNF was
\SI{1}{\nano\meter}, whereas the dielectric constant of the GNF was described by a model presented
in \cite{ytgp18JOSAB}. In order to make the GNF system share the same resonance frequency as the bowtie antenna system. The Fermi energy of graphene is chosen as \SI{0.52}{\electronvolt}, relaxation
time of \SI{0.1}{\pico\second}, and temperature of \SI{300}{\kelvin}.
Fig.~\ref{fig:PhyMechAnalog}(a) show the absorption of an isolated GNF, an isolated bowtie
antenna, and the GNF and bowtie in the bowtie-GNF hybrid system. Moreover, the corresponding field
distributions of these systems at the resonance frequency in \figref{fig:PhyMechAnalog}(a) are
given in Figs.~\ref{fig:PhyMechAnalog}(c)-\ref{fig:PhyMechAnalog}(d).

These results clearly validate our premise, namely that the influence of the GNF on the bowtie
nanoantenna is negligible. Thus, \figref{fig:PhyMechAnalog}(a) shows that there is practically a
complete overlap between the absorption spectrum of the isolated bowtie and that of the bowtie in
the bowtie-GNF hybrid system, which proves that the presence of the GNF does not affect the bowtie
nanoantenna. Moreover, the field profiles show that, at resonance, the electric field created
inside the bowtie nanoantenna gap is up to $1000\times$ larger than the field around an isolated
GNF. In addition, the field profiles in Figs.~\ref{fig:PhyMechAnalog}(b) and
\ref{fig:PhyMechAnalog}(d) suggest that the fields in the isolated bowtie nanoantenna and the
bowtie-GNF hybrid system are practically identical. This demonstrates that the scattering field of
the GNF is negligible as compared to the field in the bowtie-GNF system.

\section{Results and Discussion}\label{sec:ResDis}
\subsection{Quantum plasmons of graphene nanoflakes}\label{sec:QauntPla}
We considered first the GNFs described in the preceding section and used the TDDFT method to
investigate their optical spectra. More specifically, we used the Octopus-TDDFT code package
\cite{mcb2003CPC}, and employed the generalized gradient approximation with the
Perdew-Burke-Ernzerhof parametrization \cite{pbe96PRL}. The GNF is freestanding and it only
interacts with an external time-dependent and spatially constant electric field. We assumed that
the time dependence of the field was described by a delta-function and it was $x$-polarized. We
have performed these calculations for an undoped GNF and for GNFs with charge doping concentrations
of \SI{7}{\percent} and \SI{15}{\percent}. Here, the charge doping concentration is defined as the
ratio of the number of excess charges to the number of carbon atoms in the GNF. The computations
were performed on a computer platform containing
Intel\textsuperscript{\textregistered}~Xeon\textsuperscript{\textregistered} E5-2640v3 CPUs and 4
cores (4~GB RAM per core) were used in a generic simulation. Each TDDFT simulation required
1.04~GB of memory and was performed in about 50 hours.

\begin{figure}[!t]
  \centering
  \includegraphics[width=9 cm]{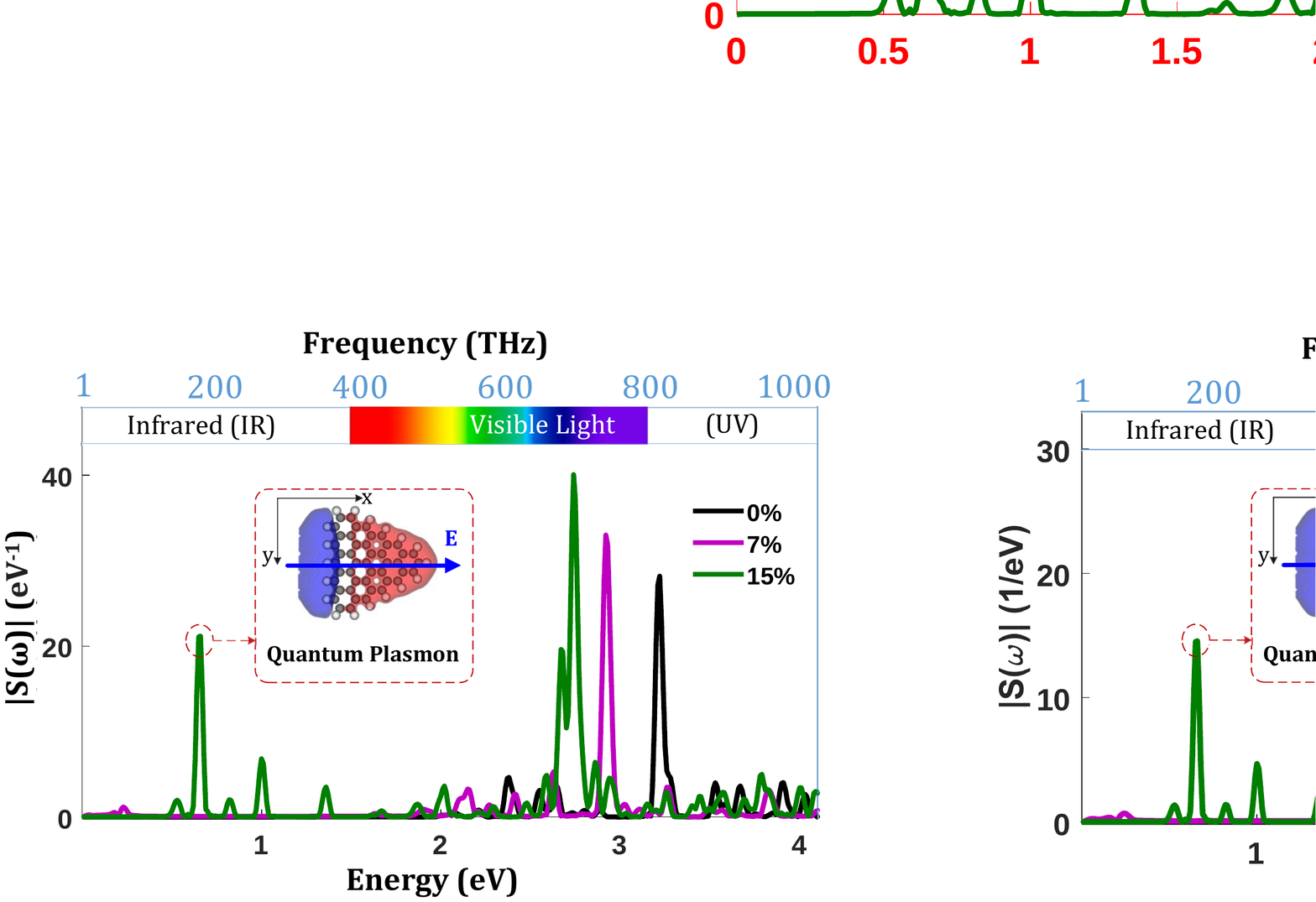}
\caption{Quantum simulation: Absorption spectra of a
graphene nanoflake calculated for several values of the charge doping concentration. In inset, the
charge distribution determined at the plasmon resonance frequency.} \label{fig:QuanPlas}
\end{figure}

The quantum response of the GNF is quantified by the dipole strength function,
\begin{equation}\label{eq:DipoleFunc}
S_{ij}(\omega)=2\omega\verb"Im"[\alpha_{ij}(\omega)]/\pi
\end{equation}
where the polarizability $\alpha_{ij}(\omega)$ is defined as:
\begin{equation}\label{eq:AlphaFunc}
\alpha_{ij}(\omega) = \frac{1}{\vert E_{j}\vert}\int[p_{i}(t)-p_{i}(0)]e^{-i\omega t}dt,
\end{equation}
where $i,j=x,y,z$ and $\vert E_{j}\vert$ is the amplitude of the $j$th component of the external
electric field. Moreover, the dynamical dipole moment $p_{i}(t)$ is evaluated as:
\begin{equation}\label{eq:DipMom}
p_{i}(t) = \int\rho(\mathbf{r},t)r_{i}d\mathbf{r}.
\end{equation}

This approach has been used to calculate the absorption spectrum of the triangular GNF, the
corresponding results being presented in \figref{fig:QuanPlas}. It can be seen in this figure that
the main resonance peak of the undoped GNF is located in the ultraviolet region, whereas when the
charge doping concentration is increased to \SI{15}{\percent} another resonance peak is formed in
the infra-red region. Moreover, the spectra presented in \figref{fig:QuanPlas} show that the
plasmon frequency decreases as the charge doping concentration increases, whereas the amplitude of
the peaks increases with the increase of the doping concentration. More importantly, new resonance
peaks emerge when the charge doping concentration increases. In order to gain more physical
insights into the nature of these resonances, we have calculated the distribution of the net charge
density at the resonance frequencies, as compared to that in the ground state. In the inset of
\figref{fig:QuanPlas}, we plot the corresponding results, determined for the GNF with
\SI{15}{\percent} doping concentration. The blue and red colors correspond to the negative and
positive net charge density, respectively. This net charge distribution does prove that this
resonance peak corresponds to collective electron density oscillations, \textit{i.e.} it can be
viewed as a quantum plasmon.

\subsection{Classical plasmons of gold bowtie}\label{sec:ClassPla}
\begin{figure}[!t]
\centering
\includegraphics[width=9 cm]{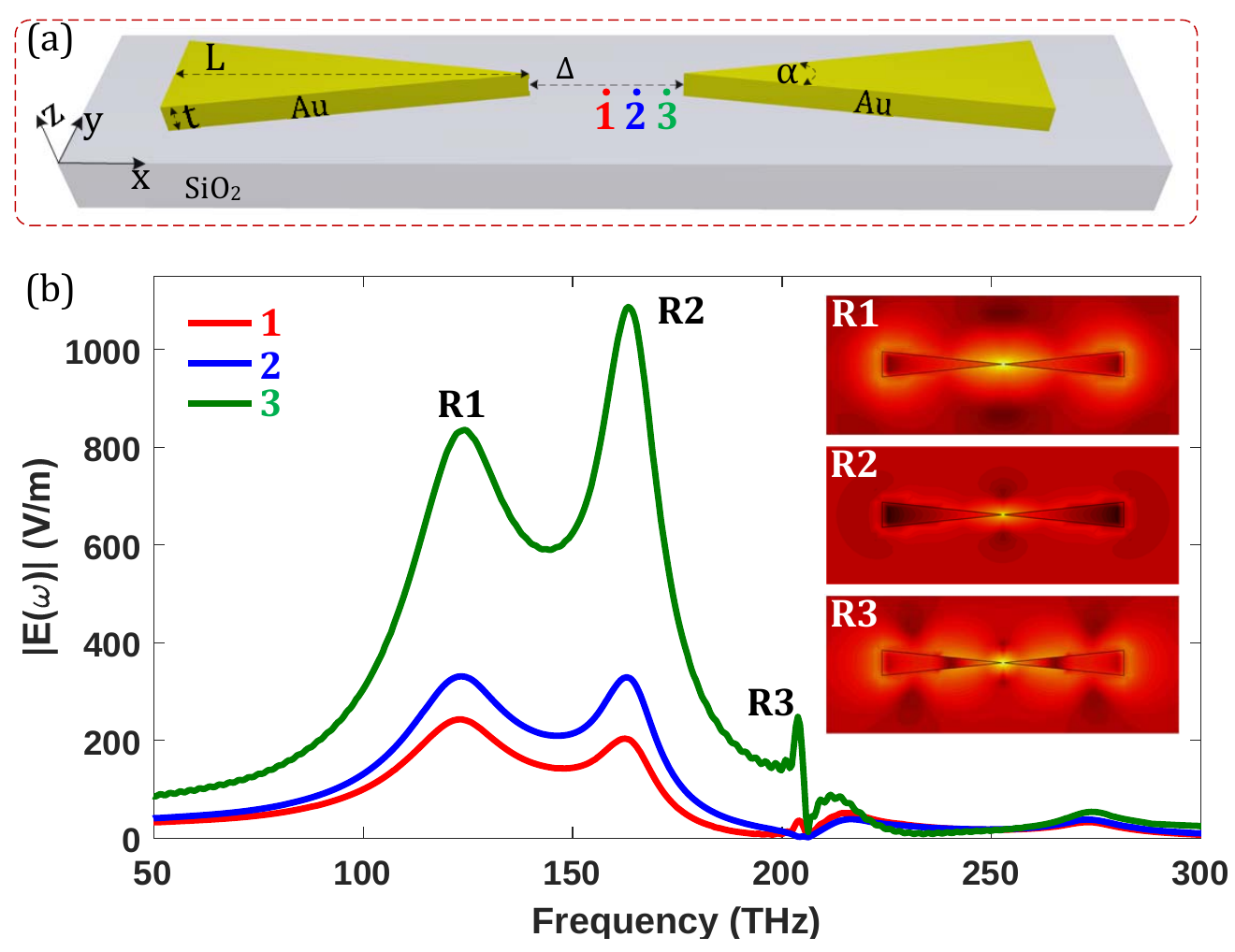}
\caption{(a) Configuration of
a gold bowtie nanoantenna and the locations where the optical near-field is probed. (b) The optical
spectra corresponding to the locations indicated in panel (a). The insets show the profiles of the
amplitude of the electric field, calculated at the resonances $R_{1}$, $R_{2}$, and $R_{3}$.}
\label{fig:ClassPlas}
\end{figure}

The second part of our study consists of the calculation of the optical spectra of the gold bowtie
nanoantenna, which is schematically illustrated in \figref{fig:ClassPlas}(a). The nanoantenna is
illuminated by a normally incident plane wave with the frequency ranging from
\SIrange{50}{300}{\tera\hertz} and the electric field is polarized along the $x$-axis. In order to
determine the optical spectrum of the nanoantenna, we computed the electric field at certain
locations in the gap using the FDTD method and then Fourier transformed it to the frequency domain.
The computations were performed on a desktop computer with a quad-core
Intel\textsuperscript{\textregistered}~Core\textsuperscript{\texttrademark} i7-4790 CPU with 8~GB
RAM per core. One such simulation required about 110~MB of memory and was performed in about 42
minutes.

The results of these calculations, determined for a nanoantenna with $L=\SI{500}{\nano\meter}$, are
summarized in \figref{fig:ClassPlas}(b). For a better comparison among the spectra, we have
normalized the spectra $|E(\omega)|$ to the amplitude of the incident electric field. The spectra
presented in \figref{fig:ClassPlas}(b) reveal several important features. First, all spectra
possess a series of resonances, the corresponding resonance frequencies being the same for all
spectra. Second, the amplitude of the optical near-field is enhanced by more than two orders of
magnitude at some particular resonance frequencies, with the enhancement factor varying with the
probing point.

In order to gain more physical insights into the characteristics of these resonances, we have
calculated the electric field profiles corresponding to the frequencies of the main spectral peaks.
The resulting plots, shown as insets in \figref{fig:ClassPlas}(b), reveal several important
findings: the resonance marked with $R_1$ represents the fundamental plasmon of the entire bowtie
nanoantenna, the resonance $R_2$ corresponds to a strongly localized plasmon formed in the narrow
gap of the nanoantenna (a so-called hot-spot plasmon), and the resonance $R_3$ represents the
second-order plasmon of the bowtie nanoantenna. Note that the resonance wavelength of the hot-spot
plasmon depends only on the shape of the plasmonic cavity and dielectric environment of the
plasmonic cavity. Moreover, particularly relevant to our study is the fact that the hot-spot
plasmon is strongly confined in the gap of the bowtie nanoantenna, which would lead to a strong
field overlap and implicitly enhanced interaction with the quantum plasmon of a GNF placed in the
gap.

\begin{figure}[!t]
\centering
\includegraphics[width=9 cm]{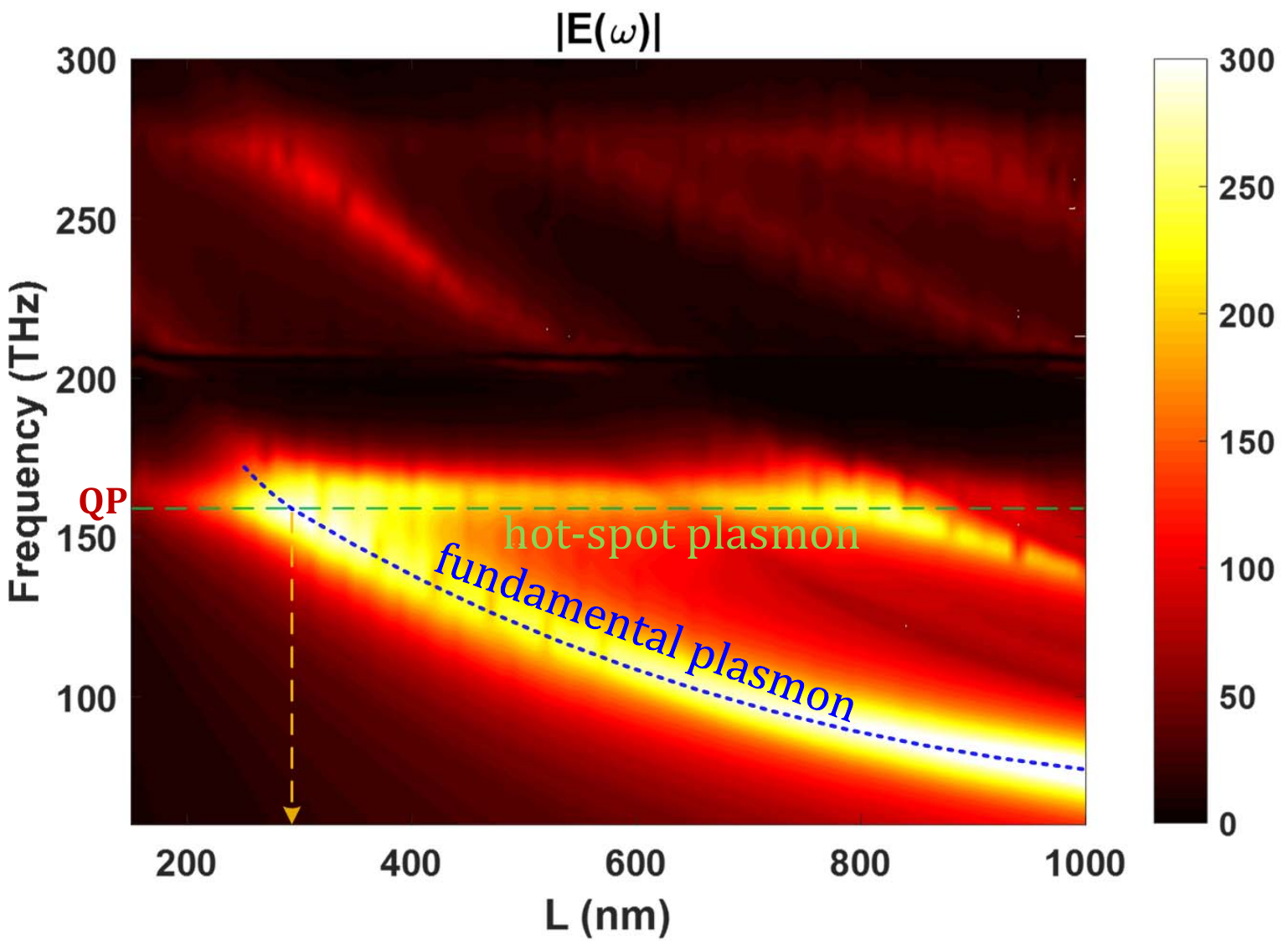}
\caption{Dispersion map of the field enhancement spectra of the gold bowtie nanoantenna, calculated at the probe
position ``1'' shown in \figref{fig:ClassPlas}(a).} \label{fig:ScanL}
\end{figure}

An enhanced interaction between the quantum plasmon of the GNF and the hot-spot plasmon of the
bowtie nanoantenna is achieved if these two plasmons have the same frequency. Therefore, in order
to reach this strongly enhanced interaction regime, we optimized the geometric structure of the
nanoantenna by varying the length $L$ so that the frequencies of these two resonances coincide.
More specifically, we scanned $L$ from \SIrange{100}{1000}{\nano\meter} while keeping constant the
values of all the other simulation parameters. The dispersion map of the corresponding field
enhancement spectra, calculated at the probe position "1" in \figref{fig:ClassPlas}(a), is given in
\figref{fig:ScanL}.

The dispersion map of the field enhancement possesses a series of bands, which correspond to
different plasmons of the system. Of all these bands, two are particularly important for our study.
The first is the plasmon band whose corresponding plasmon frequency does not depend on $L$. This
band corresponds to the hot-spot plasmon as the frequency of this plasmon depends only on the
electromagnetic environment around the narrow gap of the bowtie nanoantenna. Moreover, there is a
second plasmon band whose plasmon frequency varies with $L$. This band corresponds to the
fundamental plasmon of the nanoantenna. Moreover, the hot-spot plasmon band and the fundamental
plasmon band cross at $L=\SI{280}{\nano\meter}$, the corresponding frequency being
\SI{159}{\tera\hertz}. Importantly, this frequency is equal to the frequency of the quantum plasmon
of the GNF with \SI{15}{\percent} doping concentration. This means that for
$L=\SI{280}{\nano\meter}$ the quantum plasmon of the GNF, the hot-spot plasmon, and the fundamental
plasmon of the nanoantenna have practically the same frequency. We expect, therefore, to observe an
enhanced interaction among these plasmons for this configuration of the hybrid plasmonic system.
\begin{figure}[!t]
  \centering
  \includegraphics[width=9 cm]{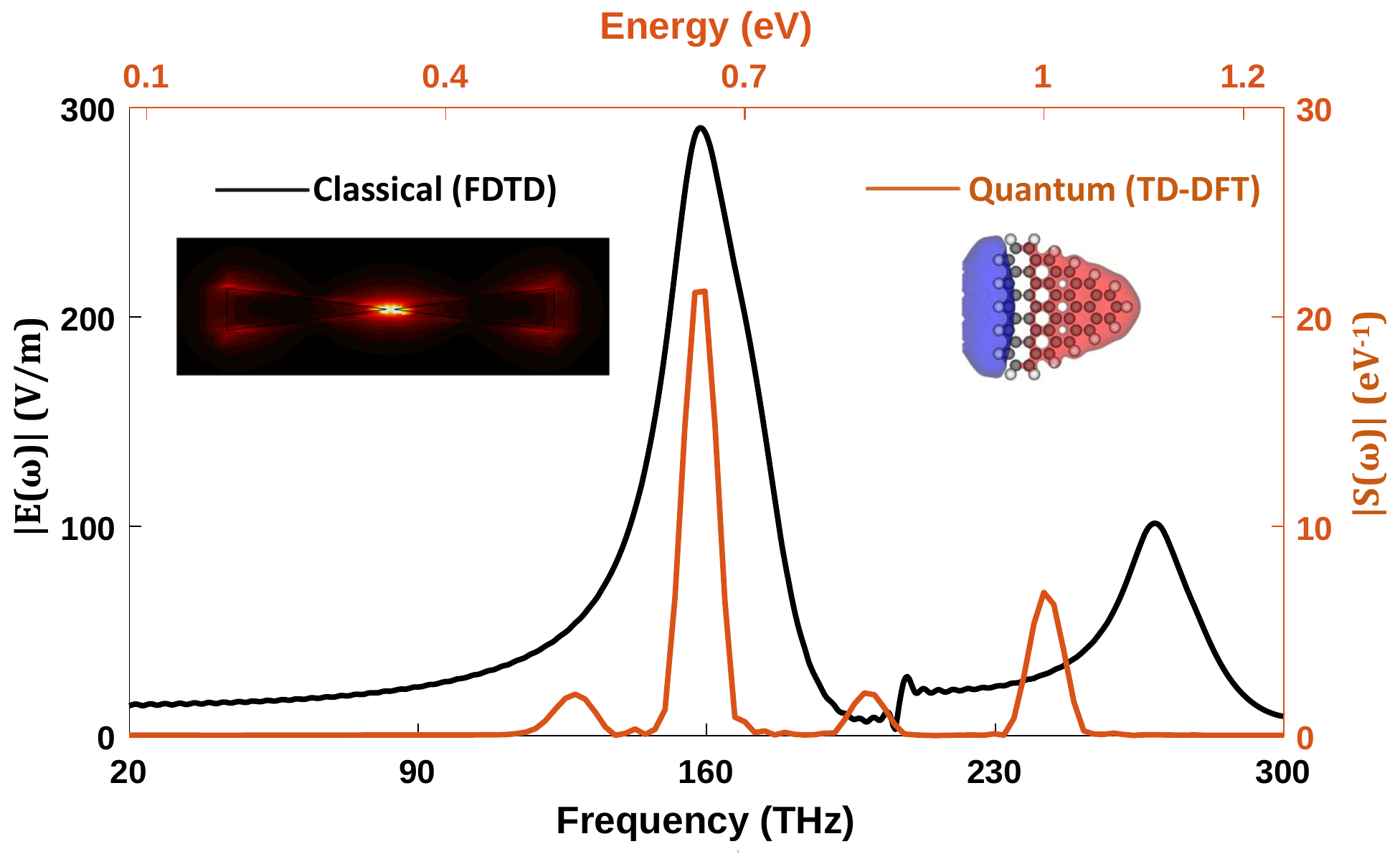}
\caption{Spectra of the bowtie
nanoantenna with $L=\SI{280}{\nano\meter}$ (black line) and GNF with a doping concentration of
\SI{15}{\percent} (red line). In insets, the profile of the optical near-field of the hot-spot
plasmon and the charge distribution of the quantum plasmon, calculated at the resonance frequency.}
\label{fig:AlignClassQaun}
\end{figure}

\subsection{Interaction between classical and quantum plasmons}\label{sec:ClassQuanPla}
The interaction between the classical and quantum plasmons can be characterized quantitatively by
analyzing the combined hybrid plasmonic system. In order to illustrate the fact that the two types
of plasmons have the same frequency when $L=\SI{280}{\nano\meter}$, we present in
\figref{fig:AlignClassQaun} the corresponding spectrum of the bowtie nanoantenna and the spectrum
of the GNF corresponding to a doping concentration of \SI{15}{\percent}. These spectra clearly show
that the two plasmon resonances are located at the same frequency of \SI{159}{\tera\hertz}. In
addition, the profile of the optical near-field of the hot-spot plasmon and the charge distribution
of the quantum plasmon, calculated at the resonance frequency, further support the plasmonic nature
of these resonances. Importantly, the spectrum of the bowtie nanoantenna suggests that at the
resonance frequency the field amplitude in the gap of the nanoantenna can be enhanced by more than
two orders of magnitude.
\begin{figure}[t!]
\centering\includegraphics[width=9 cm]{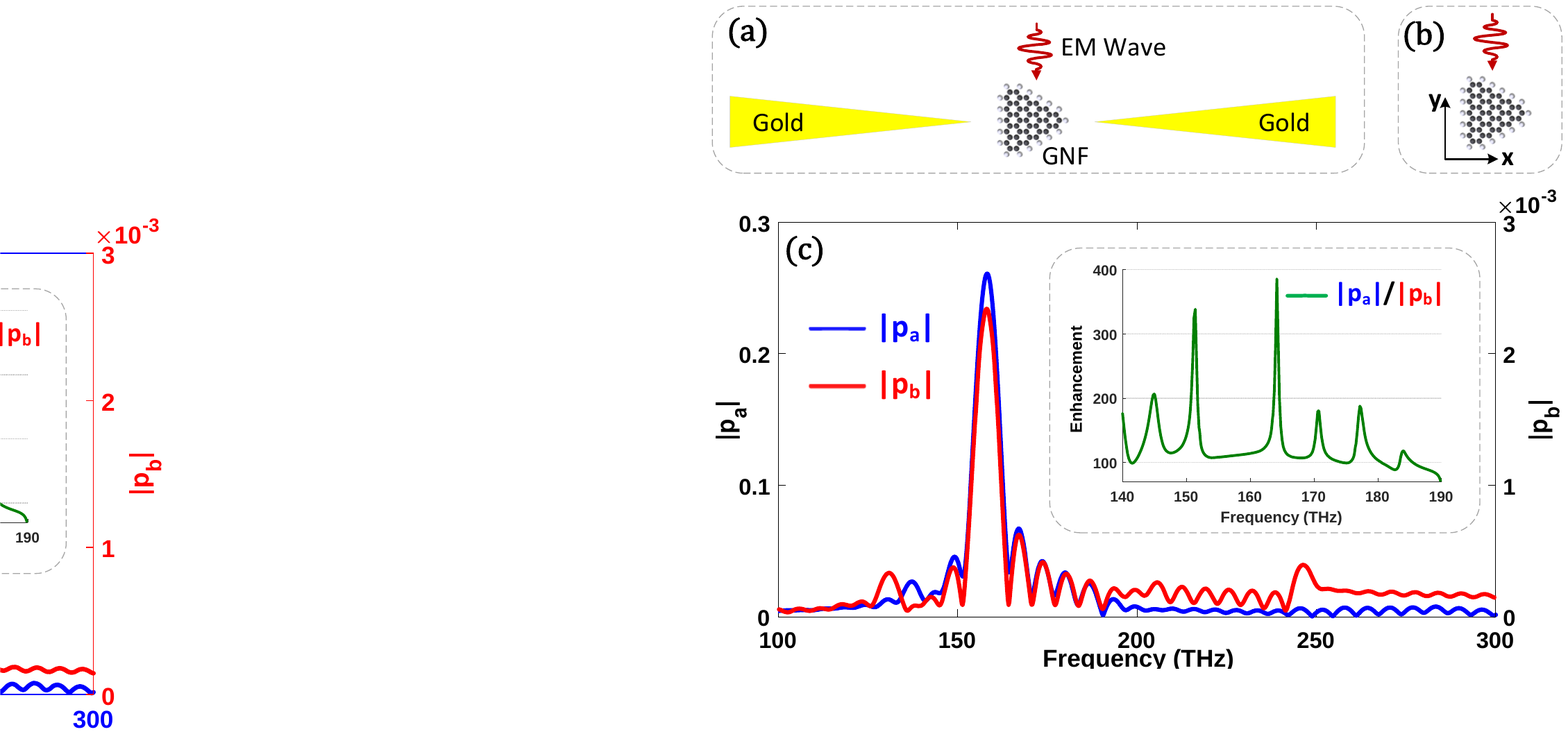} \caption{(a), (b) Schematics of a
GNF placed in the gap of a gold bowtie nanoantenna and of an isolated GNF, respectively. (c) Spectra
of the dipole moments $|p_a|$ and $|p_b|$ corresponding to the configurations shown in panels Fig.6(a)
and 6(b), respectively, and calculated under the same plane wave excitation conditions. In inset, the
spectrum on the enhancement of the quantum plasmon response defined as the ratio $|p_a|/|p_b|$.}
\label{fig:QuanEnhance}
\end{figure}

In order to assess the influence of this remarkable field enhancement effect induced by the
hot-spot plasmon resonance of bowtie nanoantenna on the quantum plasmon of the GNF, we place such a
GNF at the center of the gap of the optimized bowtie nanoantenna, as illustrated in
\figref{fig:QuanEnhance}(a). Then, using the hybrid FDTD--TDDFT method, we determined the spectrum
of the GNF. As discussed, this is performed in two steps. First, we run an FDTD computation to
determine the time-dependent electric field at the location of the GNF under a pulsed plane wave
excitation whose frequency ranges from \SIrange{100}{300}{\tera\hertz}. Then, in the second step,
we used this field as external excitation of the GNF in an TDDFT simulation. For reference, we
also calculated the spectrum of the GNF without the nanoantenna but under the same pulsed plane
wave excitation conditions, as depicted in \figref{fig:QuanEnhance}(b). The computations were
performed on a computer platform containing
Intel\textsuperscript{\textregistered}~Xeon\textsuperscript{\textregistered} E5-2640v3 CPUs and 4
cores (4~GB RAM per core) were used in a generic simulation. One such FDTD--TDDFT simulation of
the classical-quantum hybrid system shown in \figref{fig:QuanEnhance}(a) required about 1.16~GB of
memory and wall-clock time of 23.5 days, whereas a TDDFT simulation of the quantum system
presented in \figref{fig:QuanEnhance}(b) required 1.04~GB of memory and 6.8 days wall-clock time.

The results of this computational analysis are summarized in \figref{fig:QuanEnhance}. In
Figs.~\ref{fig:QuanEnhance}(a) and \ref{fig:QuanEnhance}(b), we present the two system
configurations, whereas in \figref{fig:QuanEnhance}(c) we show the spectra of the amplitude of the
dipole moments, $|p_a|$ and $|p_b|$, corresponding to the two systems. For a better comparison, we
plot in the inset of \figref{fig:QuanEnhance}(c) the ratio $|p_a|/|p_b|$, which can be viewed as
the parameter that best quantifies the enhancement of the quantum plasmon excitation. The two
spectra suggest that the frequency of the quantum plasmon of the GNF is not affected by the
excitation of the hot-spot plasmon, which means that, as expected, the effects of the quantum
plasmon on the classical one are negligible. More importantly, however, it can be seen that the
response of the quantum plasmon of the GNF can be enhanced by more than two orders of magnitude
upon interaction with the hot-spot plasmon of a specially designed gold bowtie nanoantenna.

\section{Conclusions}\label{sec:Summary}
In summary, we have applied a hybrid FDTD--TDDFT approach to study the interaction between
classical and quantum plasmons of a multiscale and multiphysical system, where a molecular-scale
graphene nanoflake is placed in the gap of a gold bowtie nanoantenna. The TDDFT simulation of the
graphene nanoflake shows that it possesses a quantum plasmon in the infrared regime if the graphene
is doped with some excess charges. We also demonstrated that the excitation of this quantum plasmon
can be significantly enhanced if the graphene nanoflake is placed inside the narrow gap of a
specially designed gold bowtie nanoantenna that possesses a hot-spot gap plasmon resonance at the
same frequency as that of the quantum plasmon of the graphene nanoflake. In particular, by
combining FDTD simulations of the bowtie nanoantenna and TDDFT calculations of the optical
response of the graphene nanoflake, we show that in the presence of the bowtie nanoantenna the
quantum plasmon response of the graphene nanoflake can be enhanced by more than two orders of
magnitude.

\section*{Acknowledgments}
The authors acknowledge the use of the UCL Legion High Performance Computing Facility
(\verb"Legion@UCL"), and associated support services, in the completion of this work.

\ifCLASSOPTIONcaptionsoff
  \newpage
\fi

\end{document}